\documentclass[a4paper,aps,prb,showpacs,twocolumn]{revtex4}
\usepackage{amsmath,amssymb,mathrsfs}
\usepackage[pdftex]{graphicx}
\usepackage[dvipsnames]{xcolor}
\usepackage[normalem]{ulem}

\begin{document}
\bibliographystyle{unsrt}

\title{ARPES in strongly disordered systems – theory of electronic band’s melting}

\author{Piotr Chudzinski}
\affiliation{School of Mathematics and Physics, Queen's University of Belfast}
\affiliation{Institute of Fundamental Technological Research, Polish Academy of Sciences, Pawinskiego 5B, 02-105 Warsaw, Poland}
\author{Lenart Dudy}
\affiliation{TEMPO Beamline, Synchrotron SOLEIL, L'Orme des Merisiers Saint-Aubin, B.P.48, 91192 Gif-sur-Yvette, France}
\affiliation{Physikalisches Institut and W\"urzburg-Dresden Cluster of Excellence ct.qmat, Universit\"at W\"urzburg, D-97074 W\"urzburg, Germany}

\date{\today}

\begin{abstract}
It is well known that translational symmetry-breaking disorder will disrupt ARPES spectra up to the point where they become invisible. However, a theoretical framework to capture this phenomenon has been largely missing. Here, based on a rigorous theory of the ARPES process, we provide this much-needed framework. In particular, we show how the frequently used sudden electron approximation has to be modified in this situation. Our main result is an argument that links the photoemission line broadening with an operator content of a disorder operator and so with the criticality of the corresponding order-disorder phase transition. For concreteness, we focus here on the frustrated 2D trigonal (pseudo-)spin model, with Ising order-disorder operators behind the transition. Still, our formalism is general and can be applied in a much broader context. 
\end{abstract}
\maketitle

\section{Introduction}

Angle-Resolved Photoemission Spectroscopy (ARPES) is one of the most important methods to study the electronic properties of materials. Its ability to probe at the same time the energy and momentum of electrons has granted the method a primary role in contemporary experimental solid-state physics. Indeed, over the last few decades, it has led us to some of the most important discoveries in the field \cite{Ding1996Pseudogap,Loeser1996Pseudogap,Im2008KondoResonance, Xia2009TI, Zhang2014MoSe2,BJ2008Iridate, Chul-Hee2019TaPWeyl, Denlinger2022}. The method is defined in reciprocal momentum space. Although it can cope with broken translational invariance in the direction perpendicular to the surface, its ability to deliver information diminishes when in-plane translational invariance is severely broken, for instance, in disordered systems. This naturally leads to a question: can ARPES be useful at all in the case of a strongly disordered system? This is a very pertinent question, especially considering recent developments in the fields such as many-body localization (MBL)\cite{ALET2018498, RevModPhys.91.021001} or dirty d-wave superconductors\cite{Chamon-dwave}. The MBL plays a particularly important role in contemporary solid-state physics. Recently given exact proof of its existence\cite{BASKO-MBL, Gornyi-MBL} inspired many numerical and experimental papers including experimental studies that showed both its realization\cite{schreiber-observation, Gross-exploring} and the slowdown of electron propagation in its vicinity\cite{Luschen-slowdown}. The disordered systems thus entered the forefront of experimental materials research. In this work, we would like to provide the theoretical framework to describe an outcome of ARPES experiment in the situation where the band structure is gradually diminished with the increasingly strong disorder. 

For simplicity, we will consider a situation of two coexisting phases on a surface of a material. In one of these phases, the carriers can easily propagate. This is our dominant “parent” state, possibly the system's ground state. The second phase provides us with a strong disorder. Within this phase, the mobile carriers are trapped. The two phases are nearly orthogonal, i.e., the tunneling between them is the slowest process during carriers propagation through the sample. Moreover, this implies that the growth/recombination of one phase into another is also prolonged. One should note that this is a realization of a strong disorder, contrary to weak scattering on impurities which may be tackled employing an additional self-energy\cite{Hwang-Born} e.g., in Born approximation. Here the large disorder strength invalidates these perturbative approximations. Since the two phases are drastically different and impenetrable one also cannot use any method of effective averaging medium\cite{Durham_1981, Bansil_1983}. The aim of this work is to derive formalism that will capture diminishing coherent band amplitude as the disorder strength increases.  

The outline of this work is as follows. First, in Sec.~\ref{Sec:Model} we introduce a specific model where our phenomenology can be realized. In Sec.~\ref{Sec:ARPEStheory}, we critically analyze the canonical theory of ARPES to extract elements that need to be re-defined. In Sec.~\ref{Sec:HTphase}, we derive formulas for ARPES spectra in the presence of disorder. Finally, in Sec.~\ref{Sec:Discussion}, we provide a broader context of our study and give examples of other possible realizations.

\section{Model}\label{Sec:Model}
\begin{figure}[htb]
\begin{center}
\includegraphics[width=0.70\linewidth]{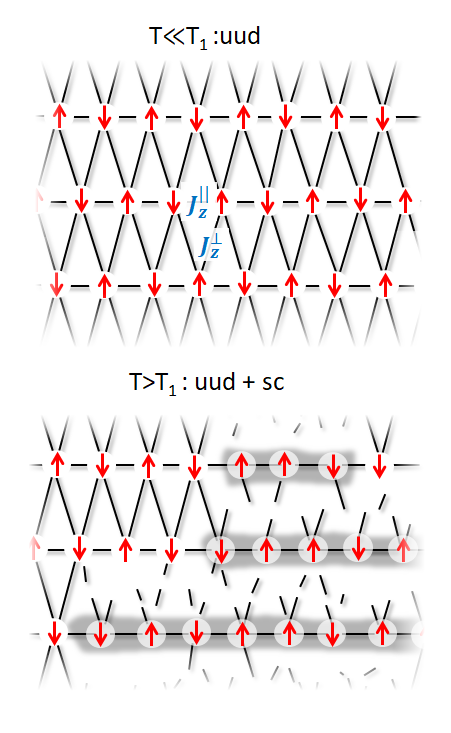}
\caption{Possible realization of our model for strong disorder. At low temperature there is a "clean" system which host a well defined bands -- dispersive energy states of pseudo-spins. At higher temperature (bottom panel) the coherence along $a$-axis is gradually lost and 1D regions that appears are able to trap photo-holes impeding their immediate propagation after the photo-emission event.}
\label{Fig:Model}
\end{center}
\end{figure}
\subsection{The idea of dimensional cross-over}

As mentioned in the introduction the idea is to introduce some external control parameter, that will govern the strength of the disorder, such that one can gradually ramp it up. The most natural candidates would be either surface irradiation or mechanically induced plasticity. The issue is whether, in an arbitrarily chosen material, the disorder induced by these two tools fulfills other criteria mentioned in the introduction. We then decided to focus on a dimensional cross-over between two insulating phases. 

To convince the reader that the peculiar situation of two impenetrable, topologically incompatible phases, can be realized, we shall focus on a strongly anisotropic 2D triangular anti-ferromagnet. In this system, there are two possible phases\cite{Starykh2015}: 1) the 2D phase, later called the uud phase, where natural excitation are vertexes on each triangle\cite{Lecheminat-Z3}, and 2) the quasi-1D phase where natural excitations are solitons\cite{Essler-1DMott}. Since the natural excitations are topologically incompatible, it is impossible that one phase will easily recombine into another. Moreover, from the theory of dimensional crossover, we know that, when perpendicular coupling $V_{\perp}$ depends on an internal degree of freedom, by increasing temperature we should reduce the perpendicular coherence and increase the amount of the secondary 1D phase. Hence, by creating a thermal mixed state, one increases the amount of disorder in the system. We see that this offers one practical platform to realize our theoretical model, and so, for concreteness, we shall stick to it in this work.

\subsection{The Hamiltonian}

ARPES measures the propagation of fermions and so we need a model where a full fermion is present. We consider a 2D surface described by Hubbard-type $U-V$ model on a 2D triangular lattice: 

\begin{widetext}
\begin{equation}
{{\cal H}_{eff}}=\sum \limits_{k \in \textrm{1$^{st.}$BZ}}\widetilde {\varepsilon \left( k \right)} \widetilde {c_k^ + }\widetilde {c_k^{}} + \sum \limits_i U{n_{m,i}}{n_{m,i}} +\sum \limits_{i,j \in  \,\textrm{chain}\;m} V_{\parallel }\left( j \right){n_{m,i}}{n_{m,i + j}}+ \sum \limits_{i,m} {V_ \bot }{n_{m,i}}{n_{m \pm 1,i \pm 1/2}} +  \ldots 
\end{equation}
\end{widetext}

Instead of working with a single particle description $\widetilde {\varepsilon \left( k \right)}$ we now consider a strongly correlated system dominated by interactions. When $U$ is by far the largest energy scale in the problem, so we can exclude double occupancies and map the problem onto a pseudo-spin system where ${V_\bot }$ and ${V_{||}}$ take the role of an anisotropic Ising-interaction. The inter-site hybridization takes the role of an equally anisotropic Heisenberg term. A self-localized hole's occupancy (or lack of it) translates into a pseudo-spin up (or down) state.  Therefore, by taking the limit $U \to \infty $ we arrive at an effective description in terms of pseudo-spins. The 2D pseudo-spin $XXZ$ model reads:

\begin{widetext}
\begin{equation}
{\cal H}_{XXZ} = \sum \limits_{i} {J_{xy}}(S_i^ + S_{i + 1}^-  + h.c.) +  \sum \limits_{i,j \in \,\textrm{in - chain}} J_z^\parallel S_{z,i} S_{z,i + j} + \sum \limits_{i\in \textrm{in-chain},n \in \textrm{n.n\;chain}} J_z^ \bot S_{z,i} (n) S_{z,i} (n + 1)
\end{equation}
\end{widetext}

This 2D $XXZ$ model on the triangular lattice is the simplest model example of frustrated magnetism and, as such, has been a subject of many, primarily numerical studies (see Ref. \onlinecite{Starykh2015} and references therein, also Ref. \onlinecite{Starykh2010} and Ref. \onlinecite{Starykh2007}) . It is known that for a magnetic field (chemical potential) close to 1/3 the model has two phases depending on the anisotropy. One of them is the 2D uud phase and another quasi-1D phase that can be distinguished by the presence of $\times$2$\times$4 solitons. We are interested in a situation where ARPES probes a mixed state, e.g. thermal state, where both these phases coexist.

The $J_z^ \bot $ or to be more precise ${{\rm{\Delta}}_1}\sim J_z^ \bot$, the energy of screened (effective) perpendicular coupling, sets the energy scale at which perpendicular coherence is gradually lost. Then there exists a low-temperature phase where the system is in a purely uud state and at $T_1=\Delta_1$ the system is in a mixed state of the two phases.

\section {The theory of the ARPES process}\label{Sec:ARPEStheory}

A canonical way of obtaining the photo-electron intensity is to assume that it is proportional to the single-particle propagator inside the sample and compute its self-energy from the momentum-conserving interactions. 
Here we show that when retardation effects are present in the coupling between electronic liquid and light, then the spectral characteristics of this coupling will affect the photo-electron intensity. In this section we shall first revisit the general ARPES theory to tackle this situation, particularly the sudden approximation (Sec.\ref{Sec:sudden_approx}). Then, we can give a precise statement of the disorder problem (Sec.\ref{Sec:validysudden_approx}). In the following, we perform calculations for LT and HT regimes. In the first case (Sec.\ref{Sec:LTphase}), the translational invariance holds, enabling us to link to standard formalism. In the second case (Sec.\ref{Sec:HTphase}), we employ our findings to incorporate the effect of N-1 particles on the top of the single-particle propagator. 

\subsection{ARPES theory: the principle of sudden approximation}\label{Sec:sudden_approx}

\begin{figure}[bt]
\begin{center}
\includegraphics[width=0.7\linewidth]{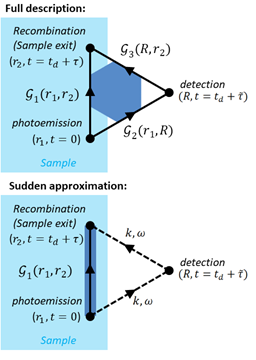}
\caption{(Upper Panel) Triangular Feynman-diagram in order to explain the photoemission process. The electron is emitted at $\left( {{r_1},t = 0} \right)$, then it goes into detector located at  $R$ and, finally at ${r_2}$ and $t = {t_d} + \tau $, it recombines with the photo-hole that was propagating through the many-body system. The overlap of these processes contributes to the signal visible at detection. (Lower Panel) Reduction of the upper diagram according to the sudden approximation. Both vertices at the detection corner are simplified by spherical harmonics with well-defined $k$ and $\omega $. Any long-distance electron-electron vertex corrections are neglected. (adapted from Ref. \onlinecite{Berthod2018})}
\label{Fig:ARPES_proc}
\end{center}
\end{figure}

The general photoemission process can be described by a triangular Feynman-diagram, see Fig. \ref{Fig:ARPES_proc} (upper panel) and compare with e.g. Ref. \onlinecite{Berthod2018,Almbladh_2006}. It consists of the photoemission corner, the detection corner, and the recombination corner at the sample exit. There are three creation and three annihilation operators and the electromagnetic field with vector potential $\vec {\cal A}\left( r \right)$ acts at the photoemission and recombination corners. Usually, one splits the six-fermion correlator into three propagators between these corners requiring altogether three Green functions (symbolized by the 
${\cal G}$'s in Fig. 1) to describe the complete process. In the most general case (e.g., to account for intrinsic and extrinsic losses), instead of the empty PES triangle (bottom panel), one has to deal with a filled triangle where there is a correlation between the propagators; For example, as the photo-electron leaves the sample, it keeps interacting with the photo-hole inside the sample. These vertex corrections are symbolized in the upper Fig.1 by the dark blue hexagon. The PES current intensity $I\left( {k,\omega } \right)$ is proportional to a quantum superposition of all these processes. Hence, a large 
$I\left( {k,\omega } \right)$ is equivalent to a constructive quantum interference established on the entire diagram. In a simplified manner, one can think of it as Einstein’s “spooky interaction on a distance” between free carriers going into the detector and a photo-hole left in the sample. Only one specific photo-hole is selected. 

To simplify and reduce the description to a single correlation function, one uses the so-called sudden approximation, see the lower panel of Fig. 1. It is based on the assumption that the photoemission process is so fast that the rest of the many-body system is unaltered when exactly one electron is excited and removed. This is equivalent to neglecting any long-distance electron-electron vertex corrections at the photoemission corner. If one neglects the same vertex corrections at the recombination corner, then the time evolution in the sample decouples; and we arrive at an expression for the photocurrent that is close to Fermi’s-golden-rule expression:

\begin{widetext}
\begin{equation}
I = \int\int d{r_1}d{r_2} \; \left\langle{\Psi_f^{N - 1} (r_2)}\right | {\vec {\cal A}(r)} \nabla_{r = r_2} \left | {\Psi_f^{N - 1}(r_2)}\right\rangle \; \left\langle{\Psi_i^{N - 1}(r_1 )}\right | {\vec {\cal A}(r ) \nabla_{r = {r_1}}}\left | {\Psi_i^{N - 1} (r_1)} \right\rangle  \label{star}
\end{equation}
\end{widetext}

where the wave-functions above are many-body wave functions, and we integrate over all possible locations of photoemission-recombination events. In the sudden approximation, one furthermore declares a simple tensor product:

\begin{equation}
{\rm{\Psi }}_i^{N - 1}\left( {{r_1}} \right) = \left| {\psi_i (r_1)} \right\rangle \otimes  \left| {N - 1}\right\rangle \label{one}    
\end{equation}

which effectively decouples the propagation of the $N - 1$ many-body system from the photo-electrons propagation. It is important to note that this single-particle wave-function ${\psi _i}\left( {{r_1}} \right)$ can only be strictly speaking defined near the space point of the instantaneous event. The ${\psi _i}\left( {{r_1}} \right)$ can be obtained either in the three-step approximation or (more accurate) in the single-step approximation.

Our description in the sample has been real space. However, we apply a monochromatic beam of photons given by the vector-potential  $\vec {\cal A}\left( {\omega ,k} \right)$ with a well-defined angle of impact, with real-space description, e.g., $\vec {\cal A}({r_1}) = {\rm{exp}}\left( {i{r_1}k} \right)\vec {\cal A}\left( {\omega ,k} \right)$. The detection counts electrons with a well-defined $\omega $ in the basis of spherical harmonics. When we further assume no interactions (no vertex corrections) on the detector and free-electron propagators out of the sample ${G_2}\left( {\omega ,k} \right) = \;{G_3}\left( {\omega ,k} \right){\rm{\;}} = {G_0}\left( {\omega ,k} \right)$  then from the ortho-normality relation of spherical harmonics 

\begin{equation}
    \int dR \, \exp{(-ik'R)} \exp{(ikR)} = \delta(k-k')
\end{equation}	
	
it follows that only that electron with the specific, well-defined momentum $k$ can be selected to give a signal at detection. This also implies that the final state at the recombination point has to be a simple tensor product 

\begin{equation}
\left| {\rm{\Psi }}_f^{N - 1}\left( {{r_2}} \right) \right\rangle = \left|{ \psi _f(r_2;k)} \right\rangle \otimes  \left| {N - 1}\right\rangle    
\end{equation}

So the final state, after propagating through detection and recombination corners, has been identified as a simple tensor product of a single particle in a state ${\psi _f}$ with a well-defined momentum k. In a system with where the translational invariance is obeyed, one can also express the initial state:
\begin{equation}
{\psi _i}\left( {{r_1}} \right) = \int dk\;{\psi _k} \left\langle k{\rm{|}}{r_1}\right\rangle\quad 
\label{doublestar}
\end{equation}

And since the action of the dipole operators  $\vec{\cal A} (r) \nabla_{r_1}$ and $\vec {\cal A}\left( r \right)\nabla_{r_2}$ is the same at every point of a translationally invariant system, one arrives at Fermi's golden rule:

\begin{equation}
I = \left| \left\langle{\psi _f(k) }\right | \vec{\cal A}(r) \nabla \left| {\psi _i ( r )} \right\rangle \right|^2 \label{doubledollar}
\end{equation}
                                                            
This formula is commonly used as a starting point for ARPES analysis. By substituting Eq.(\ref{doublestar}) into Eq.(\ref{star}), we see that only one combination of phases/amplitudes of events at various points ${r_1}$ gives the desired constructive interference on the entire diagram, the one with a well-defined momentum k. Hence we conclude that ARPES measures the single-particle spectral function $Im\left[ {{G_1}\left( {\omega ,k} \right)} \right]$ with a well-defined $k$ and $\omega$. 

At this point, it is convenient to think about one selected, distinguishable particle in the k-mode that enjoys a coherent propagation through the sample, sometimes\cite{Bardyszewski_1985} called the “blue particle”, as it does not suffer any red-shifting incoherent losses. An example of the projection procedure on the plane wave final state is performed in Sec.\ref{Sec:LTphase}.

\subsection{The validity of the sudden approximation in a disordered system} \label{Sec:validysudden_approx}

The translational invariance is clearly broken in a strongly disordered sample, and the reasoning described above, e.g., Eq.(\ref{doublestar}), does not hold anymore. Rooted in the general aspects described above, the problems of a description of ARPES in the disordered medium of uud and $\times$2$\times$4 regions are twofold. The first is to assure the validity of the sudden approximation as much too frequently\cite{Almbladh_2006} all the steps leading to Eq.(\ref{doubledollar}) are assumed to be straightforward implications of the sudden approximation. The second one is to find a way to compute ${G_1}$ in the situation where the system's response to incoming photons depends on the exact place where the emission event took place. In particular, since the assumption still holds, we need to find a way to form a constructive interference with the spherical-harmonically single-electron wave-functions $\left|{\psi _f}\left( {{r_2};k} \right)\right\rangle$. There are two equivalent ways of tackling our problem: \\

(1)	We take the basis where the ARPES triangle-diagram in Fig.\ref{Fig:ARPES_proc} is empty. The photo-electron is suddenly removed from the sample, and we can safely assume that it does not interact (no long-distance retardation effects) with a photo-hole left inside the point  ${r_1}$ of the sample (so Eq.(\ref{one}) holds). Photo-hole acquires non-local self energy, an attempt to compute this quantity has been given in Ref.\onlinecite{Yasuchara-weird-S} but in a simpler translationally invariant case. Within the disordered sample, we need to consider a time evolution of an object that depends strongly on the position where the photoemission event took place. While Eq. (\ref{star}) holds, we do not know how to perform the real-space integral, and then, Eq. (\ref{doubledollar}) does not hold. The $G_1(r_1,r_2)$ becomes a real-space-only object that would be extremely tedious to time-evolve along all possible paths, all possible realizations of disorder. The analytical progress of such a theory is hard, and here we will not follow this route.  However, this picture proves that it is possible to construct the so-called “blue particle” of the sudden approximation, albeit very complicated and dressed up with a time-evolving disorder. \\

The way (1) above would be the only way to proceed for a massively disordered system with no extra information at hand. However, in our case, we know that in the uud zones, which percolate (or are close to percolation) at least at the beginning of the HT-phase, the momentum remains a good quantum number. This brings us to the second way\\

(2)	We can then divide  $G_1({r_1},{r_2})$ into two components, an effective $\widetilde {{G_1}}$ inside a uud zone and the rest of the time evolution in the $\times$2$\times$4 regions. To apply Eq.(\ref{doublestar}) (valid for the uud part of evolution), the vertex corrections (that describe a time-evolution inside $\times$2$\times$4 region, immediately after the photoelectron escape) have to be located in either the photoemission or the recombination corner of the triangle PES diagram. As shown below, these corrections capture the contribution in the photo-electron intensity describing the path a photo-hole has to endure in the disordered medium of the $\times$2$\times$4 regions. Hence, we work with a particle that has lost some of its energy to propagate out of the disordered zone. This slightly red-shifted particle still follows a coherent time-propagation, and, thereby, it is distinguishable from the incoherent background\cite{Bardyszewski_1985,McMullen_1976} - by analogy, one can call it a “green particle”.

Adding the above vertex corrections is similar to incorporating an extra “fourth-step” in the three-step model. This extra step involves all N-1 particles in the Hilbert space undergoing a coherent rotation. In this Dirac picture, the  $\widetilde {{G_1}}$ and vertex corrections are separable as Heisenberg and Schrodinger time evolutions. The procedure is applied in Sec. \ref{Sec:HTphase}.

\subsection{Description in the translationally invariant medium (LT phase)}\label{Sec:LTphase}

For illustrative purposes, let us begin with a description at $T = 0$. The state of the system before the transition is translationally invariant as it consists entirely of the  $\left| {uud} \right\rangle$  sites on all sites:

\begin{equation}
{\left| {{\psi _0}} \right\rangle^{T = 0}} = \mathop \prod \limits_i^N {\left| {uud} \right\rangle^{\left( i \right)}}
\end{equation}

Now, we include the action of a photon (an electromagnetic field $A\left( t \right)$), which leads to a final state where a photo-electron has been suddenly removed. The coherent band we consider, called $S_{coh}$ in the following, is a surface spectral feature. Since carriers $S_{coh}$ exist on the surface therefore, there are no issues with escape length or penetration depth. The final state at $t=0$, intermediately after the photoemission event, can be expressed as a sum of Slater determinants in a real-space basis:   

\begin{equation}
{\left| {{\psi _{im}}} \right\rangle^{T = 0}} = \mathop \sum \limits_{x_0}^N {\left| {uud} \right\rangle_{x_0}^{*}} \otimes \widetilde{\prod \limits_{x \ne {x_0}}} {\left| {uud} \right\rangle^{\left(x \right)}}
\end{equation} 

Where we have singled out one particular site as the rapid photoemission event takes place at a given site  ${x_0}$ containing suddenly one photo-hole, written as $\left| {\cdots} \right\rangle_{{x_0}}^*$,  $\otimes$ is a tensor product, and  $\widetilde{\prod \limits_{x \ne {x_0}}}$ is an anti-symmetrized product (i.e., a Slater determinant over all other occupied sites).  The photo-hole now propagates through the sample (to be precise, in the strongly correlated model under consideration a full fermion is a fusion of fractional particles with charge e/3 as derived in Ref.\cite{Lecheminat-Z3}), and, upon exiting the sample, it re-combines with a photo-electron on the detector. We now make a key assumption of this reasoning: Since at $T=0$ the sample is translationally invariant (along the edge direction), hence the momentum is a good quantum number and only states with well-defined momentum emerge out of the ${x_0}$-summation. Therefore, the initial state can be written as a simple tensor product 

\begin{equation}
\left| {\psi_i} \right\rangle^{T = 0} \sim c_k^+ (0) \left| {\psi_{f - im}} \right\rangle^{T = 0}\sim c_k^+ (0) \left|{\psi _k}\right\rangle \left| {N - 1} \right\rangle_{uud}^{(0)}
\end{equation} 

where the $(0)$ subscript indicates that the $N-1$ particles stay in their unaffected quantum state. It is identical to the form of the final state that can be deduced from the propagation on the triangle diagram.

For the final state, by invoking momentum conservation (which we assume to be obeyed), we use the statement derived at the beginning of this section that only a particular combination of real space Slater determinants survives the quantum interference:

\begin{equation}
{\left| {{\psi _f}\left( t \right)} \right.^{T = 0}} =\sum \limits_s c_k^ + \left( t \right)\left| {N - 1} \right\rangle_{uud}^s
\end{equation}

where the desired combination has been generated by applying the $c_k^+$ operator acting on the $N-1$ other fermions in the uud state. The summation over $s$ is over all possible trajectories that will produce the desired state with momentum $k$,
 e.g., on its way, the photo-hole may emit/absorb a phonon, plasmon, etc. …  (when the self-energy is zero, there shall be only one term left in this sum).  The expression for the photo-electron intensity simplifies remarkably, especially if evaluated in the Heisenberg picture:

\begin{align}
    I &= \; ^{T = 0}\left\langle {\psi _f} | {\psi _i} \right\rangle^{T = 0}& \nonumber\\
    &= \sum \limits_s {}_{uud}^s \left\langle {N - 1} \right|\otimes \left\langle{\psi _k}\right|c_k (t) c_k^+ (0) \left|{\psi _k} \right\rangle \otimes \left | {N - 1}\right\rangle_{uud}^{(0)}& \nonumber\\
    &= A(k,\omega)  \sum \limits_s {}_{uud}^s \left\langle {N - 1} \right |1 \left| {N - 1}\right\rangle_{uud}^{(0)}.& \label{doublecross}
\end{align}

Where $A\left( {k,\omega } \right) = {\psi _k}|c_k^ + \left( t \right)c_k^{}\left( 0 \right)|{\psi _k}$ is the spectral function, a correlator whose expectation value can be computed on a single disentangled state of the Hilbert space since we assumed that the “blue particle” is disentangled from the remaining $N-1$ particles.  The time evolution of the single hole part of the final state (${{\cal U}^+ (t)\;c_k^+ {\cal U}(t) }$) combined with the initial state $c_k^+ (t=0)$ gives (an imaginary part of) a propagator whose temporal Fourier transform is the single-particle spectral function with a given momentum $k$. Based on this, it is usually found that ARPES measures the spectral function for carriers with given momentum as expressed by the 2nd equality on the r.h.s. As a result, our ARPES measures the excitation of the $S_{coh}$ carriers with well-defined energy and momentum. Please note that we are making here a hidden assumption that the remaining bracket is unitary: the $c_k^+ (t)$ electron has not changed any other states such that the overlap between the $N - 1$ states is perfect.

\section{Solution for the disordered medium (HT phase)}\label{Sec:HTphase}

Based on the reasoning in Sec.\ref{Sec:ARPEStheory} we are now ready to tackle the photoemission process from the disordered system.

\subsection{Carrier propagation through the mixed state}

Now we move to finite temperatures where the initial state consists of a mixture of $\left| {uud} \right\rangle$- sites and single chain 
$\left| {sc} \right\rangle$-sites (NOTE: $\times$2$\times$4 region is equivalent to the $\left| {sc} \right\rangle$). The latter ones presumably form a fractal structure – a Brownian tree\cite{Witten-Brown-tree}. The initial state reads:

\begin{equation}
    \left| {\psi_0}\right\rangle^{T > {T^*}} =  \prod \limits_i^n \left| {uud} \right\rangle^{(i)} \otimes \prod \limits_j^m  \left| {sc} \right\rangle^{(j)}
\end{equation}

Removing a photo-electron from this state is a non-trivial operation as the state is intrinsically non-homogeneous. The translational invariance is not obeyed; we cannot assume that the momentum is a good quantum number. For the intermediate state at $t=0$, intermediately after the photoemission event, we need to write a double sum to account for the fact that an empty site can be either within $\left| {uud} \right\rangle$ or $\left| {sc} \right\rangle$:

\begin{widetext}	
\begin{equation}
    {\left| {{\psi _{im}}} \right\rangle^{T > {T^*}}} = \mathop \sum \limits_{{x_0}}^n \left| {uud} \right._{{x_0}}^* \otimes \widetilde{ \prod \limits_{\begin{array}{*{20}{c}}{x \in n}\\{x \ne {x_0}}\end{array}} } \left| {uud} \right\rangle^{(x)} +  \sum \limits_{{x_0}}^m \left| {sc} \right\rangle_{{x_0}}^* \otimes  \widetilde{\prod \limits_{\begin{array}{*{20}{c}}
{x \in m}\\
{x \ne {x_0}}
\end{array}}} \left| {sc} \right\rangle^{(x)}
\end{equation}
\end{widetext}

If the disordered state would be infinitely rigid (i.e., the $sc$ and $uud$ would be described by two independent quantum states), then the photo-hole on ${\left| {sc} \right\rangle^*}$-site would be forever locked and would not contribute to a $uud$ photocurrent. More precisely: a different ansatz for the $N$-body wave function should then be used. Namely, each observable would have been described by a simple sum of signals from  $\left| {uud} \right\rangle$ and disordered zones that contain the $\left| {sc} \right\rangle$-sites. At temperatures in the transition range, we would then expect a double peak structure (a narrow coherent feature with decreasing amplitude, stemming from the $\left| {uud} \right\rangle$ -sites, and a much broader increasing incoherent feature stemming from the $\;\left| {sc} \right\rangle$-sites. 

In our problem however, the photo-hole in the ${\left| {sc} \right\rangle^*}$-site can be extracted by applying the Ising-disorder operator $\mathop \prod \limits_i^{sd} {\hat \mu ^{\left( i \right)}}\left( t \right)$ sufficiently many times. This represents an additional time-evolution\cite{McMullen_1976} of the $\left| {N - 1} \right\rangle$-system compared to the case at $T = 0$ discussed above. This time evolution affects all the other states. We can incorporate in the triangle diagram as a vertex correction at photoemission corner and include this extra time evolution in the Dirac picture by taking an $sd$-long sequence of Ising-disorder operators $\mathop \prod \limits_i^{sd} {\hat \mu ^{\left( i \right)}}\left( t \right)$ for producing the additional time-dependent interaction and find the final form of the initial state: 

\begin{equation}
{\left| {{\psi _i}} \right\rangle^{T > {T^*}}} = \mathop \sum \limits_s \left |{\psi _k} \right\rangle\;\otimes{\left( {\mathop \sum \limits_{sd}^{} \mathop \prod \limits_i^{sd} {{\hat \mu }^{\left( i \right)}}\left( t \right)\left| {N - 1} \right\rangle} \right)_{uud}}	    
\end{equation}

Where the summation over $sd$ indicates various trajectories over which the $\prod \limits_i^{sd} {\hat \mu ^{(i)}}(t)$ have been applied. Once the photo-hole is in the domain of the uud states, its further propagator towards the relaxation is known, the momentum is now conserved, and we can follow the same lines as those computed above for the $T = 0$ case. By our reasoning, the time evolution consists now of two stages. Hence, its Fourier transform will be a convolution of two propagators: the $\left| {uud} \right\rangle$ propagator, constituting the $S_{coh}$ dispersion, and an additional propagator of the $\left| {N - 1} \right\rangle$-carriers due to the application of the Ising operators.


\subsection{Temperature dependence of ARPES in the disordered medium}

We then arrive at our main result that the photo-electron current in momentum-frequency space, analogous Eq.(\ref{doublecross}), is given by a convolution:

\begin{equation}
    I = A (k,\omega)\otimes \left\langle (N - 1)_{f_{ t = t_d ( T)}}  |  (N – 1)_i \right\rangle \label{eq:final-result}
\end{equation}	

where we see that standard spectral function of the coherent part $A (k,\omega)$ is convoluted with a term that describes the Dirac time evolution of remaining $N-1$ fermions. The subscript indicates that we consider a final state after a given time $t_d$, see below for its definition.

In our case the $sc\rightarrow uud$ tunneling is local, hence the momentum dependence of Dirac part drops out. Furthermore, for our immobile $sc$ units, in 1D regime it is known that even for interacting case the effect of scattering can be absorbed as a purely phase factor\cite{EggertAffleck}. In a language of Ref.\onlinecite{Braun-phononARPES} we can thus safely neglect the dynamics of amplitudes and focus on phase evolution (as in Ref.\onlinecite{Feder_1981}) of the disorder operators $\hat{\mu}$. In this case the convolution can be reduced to a simple multiplication.

To complete our analysis, we can now compute this overlap between the initial and final $\left| {N - 1} \right\rangle$  states at finite temperature, which we shall call Dirac evolution amplitude:
\begin{equation}
{A^D} = \left\langle {\left( {N - 1} \right)_{f_{t = {t_d}(T)}} | ({N - 1})_i} \right\rangle   
\end{equation}

This overlap enters into the total overlap between $ \left\langle{\psi _f}{\rm{|}}{\psi _i}\right\rangle$ as an extra, non-unitary factor. We assume that there exist a typical delay time ${t_d}$ which is needed to remove the photo-hole from the $\left| {sc} \right\rangle$  zone. Taking a constant inter-site tunneling rate, this delay is merely proportional to a typical escape length ${x_d}$ from inside the aggregate of $sc$-zones which is a fractal. From the diffusion-limited aggregation\cite{DLA-orig} we know that the fractal is a Brown tree\cite{Witten-Brown-tree}. By increasing the temperature $T$  we increase the number of $\times$2$\times$4 units (as there are more available many-body states in an energy window ${k_B}T$), the volume of the fractal ${V_d}$ (Brown tree) scales with the number of states like ${N^{{d_H}}}$ (where ${d_H}$ is the Hausdorff dimension) while the $X_d\sim{V_d}^{1/d}$. In our case ${d_H} = 2$ and $d = 2$. Hence, we deduce that $t_d\sim T$. Having this information, we assume in the following that the convolution of the Dirac evolution amplitude with the uud spectral function results in a Lorentzian-like function showing a continuous change of maximum intensity and width.

For instance, let us consider the strength of the maximum intensity of photo-electrons. When we increase the temperature above ${{\rm{T}}_1} = {\rm{\;}}{{\rm{\Delta }}_1}$ the amplitude of the maximum shall decrease as the overlap between initial and final Ising operators configuration degrades. This is because close to the transition, due to $\prod \limits_i^{sd} {\hat \mu ^{\left( i \right)}}\left( t \right)$ fluctuations, the Ising order is known\cite{Wu1976Ising} to decay like $1/{t^{1/4}}$. Hence, we expect $1/{T^{1/4}}$ decay of the photocurrent amplitude that is:
\begin{equation}\label{eq:fin}
    A_D=1/T^{1/4}
\end{equation}
simple relation that can be readily probed by experiments. It should be emphasized that the temperature dependence in Eq.\ref{eq:fin} is not due to any internal phononic-type dynamics\cite{Braun-phononARPES} on each site, but it is a many body effect of a coherence loss within the $sc$ conglomerate. In other words there is a loss of coherence between two Ising $\hat{\mu}$ operators applied at a certain interval of time $t_d$.  

\section{Discussion and Conclusions}\label{Sec:Discussion}

Firstly, within our example, we would like to point out,  that the formalism proposed here is much more general than for a transition between the uud and sc phases only induced by temperature. One can think of other external parameters inducing the transition. For the specific case of screened Coulomb interaction mediated inter-chain coupling, these can be, i.e., local chemical doping, local tension or a local screening by an external inhomogeneous electrode. This richness of possible means of control comes from the fact that the screening has an internal degrees of freedom and is a great advantage of the proposed platform based on a dimensional-cross over.    

In our example, we had certain operators that induced the disorder. These were the Ising operators because these are the ones that diagonalize the strongly-correlated Hamiltonian. An advantage of our example was that, by construction through topological protection, it created \emph{localized and stable} disordered units. Generally, one can consider other operators that induce disorder.  We could consider a variation of the chemical potential for introducing disorder. However, this approach will only succeed in cases where the Hamiltonian results in a single particle description (hence: disorder leading to Anderson localization). Then, the critical behaviour of the chemical-potential operator is known; it is a marginal operator. One can also consider a model where electrons are coupled with a local boson, e.g. an optical phonon mode. If the disorder is then proportional to the local density of phonons ($\equiv$ displacement), the criticality is proportional to the scaling operator of the phononic density.

The main outcome of this work is that it provides a formalism to quantify how an ARPES band is disappearing in the presence of disorder. Remarkably, from the energy scale of its disappearance as well as from the profile of the broadening curve, we can gather a lot of information about the nature of a phase transition induced by that disorder. The underlying reason, why the band broadens, is the breaking of the translational symmetry. This implies that the momentum is not any longer a good quantum number and that the spectral weight is broadly distributed within the reciprocal space. We have taken here a particular model which allows for only one selected band to be affected. It should be emphasized that, while Eq.(\ref{eq:final-result}) describes how the spectral weight of the \emph{coherent} part of the band disappears, it does not invalidate the ARPES-sum-rule. The ARPES sum-rule can be traced back to the anti-commutation rules for the underlying fermionic fields\cite{GursoyARPES} which, as such, still hold. It just requires to include all scattering channels of fermions, i.e., all the ways in which the \emph{incoherent} part of the propagator can be generated. In the specific case of the Ising model proposed here, that would mean including all the possible order-disorder operators' correlation functions. 

From the experimental perspective, our theory can also be useful in other ways, albeit more limited. Consider a system with an unknown amount of embedded disorder. The non-linear dependence of the amplitudes' suppression (and the coherent band's broadening) allows to extract, i.e. by exploring the derivative, small changes in disorder strength. This, in turn, allows to detremine the amplitude of the disorder. With this information, the entire ARPES spectra can be fitted using an additional correction that fully accounts for non-perturbative effects of the disorder.  

In \emph{conclusion} we derived here a formalism that extends the standard \emph{sudden-electron} approximation to cases where carriers are localized for a finite time during the photo-emission event. Then, the ARPES spectrum acquires an extra broadening factor which can be interpreted as a vertex correction. If there is any operator that links the ordered and disordered phases, we then showed that the broadening of the ARPES spectrum  mimics the dynamics of this operator.

\bibliographystyle{apsrev}

\bibliography{bib}

\begin{thebibliography}{36}
\expandafter\ifx\csname natexlab\endcsname\relax\def\natexlab#1{#1}\fi
\expandafter\ifx\csname bibnamefont\endcsname\relax
  \def\bibnamefont#1{#1}\fi
\expandafter\ifx\csname bibfnamefont\endcsname\relax
  \def\bibfnamefont#1{#1}\fi
\expandafter\ifx\csname citenamefont\endcsname\relax
  \def\citenamefont#1{#1}\fi
\expandafter\ifx\csname url\endcsname\relax
  \def\url#1{\texttt{#1}}\fi
\expandafter\ifx\csname urlprefix\endcsname\relax\def\urlprefix{URL }\fi
\providecommand{\bibinfo}[2]{#2}
\providecommand{\eprint}[2][]{\url{#2}}

\bibitem[{\citenamefont{Ding et~al.}(1996)\citenamefont{Ding, Yokoya,
  Campuzano, Takahashi, Randeria, Norman, Mochiku, Kadowaki, and
  Giapintzakis}}]{Ding1996Pseudogap}
\bibinfo{author}{\bibfnamefont{H.}~\bibnamefont{Ding}},
  \bibinfo{author}{\bibfnamefont{T.}~\bibnamefont{Yokoya}},
  \bibinfo{author}{\bibfnamefont{J.~C.} \bibnamefont{Campuzano}},
  \bibinfo{author}{\bibfnamefont{T.}~\bibnamefont{Takahashi}},
  \bibinfo{author}{\bibfnamefont{M.}~\bibnamefont{Randeria}},
  \bibinfo{author}{\bibfnamefont{M.}~\bibnamefont{Norman}},
  \bibinfo{author}{\bibfnamefont{T.}~\bibnamefont{Mochiku}},
  \bibinfo{author}{\bibfnamefont{K.}~\bibnamefont{Kadowaki}}, \bibnamefont{and}
  \bibinfo{author}{\bibfnamefont{J.}~\bibnamefont{Giapintzakis}},
  \bibinfo{journal}{Nature} \textbf{\bibinfo{volume}{382}}, \bibinfo{pages}{51}
  (\bibinfo{year}{1996}).

\bibitem[{\citenamefont{Loeser et~al.}(1996)\citenamefont{Loeser, Shen, Dessau,
  Marshall, Park, Fournier, and Kapitulnik}}]{Loeser1996Pseudogap}
\bibinfo{author}{\bibfnamefont{A.}~\bibnamefont{Loeser}},
  \bibinfo{author}{\bibfnamefont{Z.-X.} \bibnamefont{Shen}},
  \bibinfo{author}{\bibfnamefont{D.}~\bibnamefont{Dessau}},
  \bibinfo{author}{\bibfnamefont{D.}~\bibnamefont{Marshall}},
  \bibinfo{author}{\bibfnamefont{C.}~\bibnamefont{Park}},
  \bibinfo{author}{\bibfnamefont{P.}~\bibnamefont{Fournier}}, \bibnamefont{and}
  \bibinfo{author}{\bibfnamefont{A.}~\bibnamefont{Kapitulnik}},
  \bibinfo{journal}{Science} \textbf{\bibinfo{volume}{273}},
  \bibinfo{pages}{325} (\bibinfo{year}{1996}).

\bibitem[{\citenamefont{Im et~al.}(2008)\citenamefont{Im, Ito, Kim, Kimura,
  Lee, Hong, Kwon, Yasui, and Yamagami}}]{Im2008KondoResonance}
\bibinfo{author}{\bibfnamefont{H.~J.} \bibnamefont{Im}},
  \bibinfo{author}{\bibfnamefont{T.}~\bibnamefont{Ito}},
  \bibinfo{author}{\bibfnamefont{H.-D.} \bibnamefont{Kim}},
  \bibinfo{author}{\bibfnamefont{S.}~\bibnamefont{Kimura}},
  \bibinfo{author}{\bibfnamefont{K.~E.} \bibnamefont{Lee}},
  \bibinfo{author}{\bibfnamefont{J.~B.} \bibnamefont{Hong}},
  \bibinfo{author}{\bibfnamefont{Y.~S.} \bibnamefont{Kwon}},
  \bibinfo{author}{\bibfnamefont{A.}~\bibnamefont{Yasui}}, \bibnamefont{and}
  \bibinfo{author}{\bibfnamefont{H.}~\bibnamefont{Yamagami}},
  \bibinfo{journal}{Phys. Rev. Lett.} \textbf{\bibinfo{volume}{100}},
  \bibinfo{pages}{176402} (\bibinfo{year}{2008}),
  \urlprefix\url{https://link.aps.org/doi/10.1103/PhysRevLett.100.176402}.

\bibitem[{\citenamefont{Xia et~al.}(2009)\citenamefont{Xia, Qian, Hsieh, Wray,
  Pal, Lin, Bansil, Grauer, Hor, Cava et~al.}}]{Xia2009TI}
\bibinfo{author}{\bibfnamefont{Y.}~\bibnamefont{Xia}},
  \bibinfo{author}{\bibfnamefont{D.}~\bibnamefont{Qian}},
  \bibinfo{author}{\bibfnamefont{D.}~\bibnamefont{Hsieh}},
  \bibinfo{author}{\bibfnamefont{L.}~\bibnamefont{Wray}},
  \bibinfo{author}{\bibfnamefont{A.}~\bibnamefont{Pal}},
  \bibinfo{author}{\bibfnamefont{H.}~\bibnamefont{Lin}},
  \bibinfo{author}{\bibfnamefont{A.}~\bibnamefont{Bansil}},
  \bibinfo{author}{\bibfnamefont{D.}~\bibnamefont{Grauer}},
  \bibinfo{author}{\bibfnamefont{Y.~S.} \bibnamefont{Hor}},
  \bibinfo{author}{\bibfnamefont{R.~J.} \bibnamefont{Cava}},
  \bibnamefont{et~al.}, \bibinfo{journal}{Nature Physics}
  \textbf{\bibinfo{volume}{5}}, \bibinfo{pages}{398} (\bibinfo{year}{2009}),
  \urlprefix\url{https://doi.org/10.1038/nphys1274}.

\bibitem[{\citenamefont{Zhang et~al.}(2014)\citenamefont{Zhang, Chang, Zhou,
  Cui, Yan, Liu, Schmitt, Lee, Moore, Chen et~al.}}]{Zhang2014MoSe2}
\bibinfo{author}{\bibfnamefont{Y.}~\bibnamefont{Zhang}},
  \bibinfo{author}{\bibfnamefont{T.-R.} \bibnamefont{Chang}},
  \bibinfo{author}{\bibfnamefont{B.}~\bibnamefont{Zhou}},
  \bibinfo{author}{\bibfnamefont{Y.-T.} \bibnamefont{Cui}},
  \bibinfo{author}{\bibfnamefont{H.}~\bibnamefont{Yan}},
  \bibinfo{author}{\bibfnamefont{Z.}~\bibnamefont{Liu}},
  \bibinfo{author}{\bibfnamefont{F.}~\bibnamefont{Schmitt}},
  \bibinfo{author}{\bibfnamefont{J.}~\bibnamefont{Lee}},
  \bibinfo{author}{\bibfnamefont{R.}~\bibnamefont{Moore}},
  \bibinfo{author}{\bibfnamefont{Y.}~\bibnamefont{Chen}}, \bibnamefont{et~al.},
  \bibinfo{journal}{Nature Nanotechnology} \textbf{\bibinfo{volume}{9}},
  \bibinfo{pages}{111} (\bibinfo{year}{2014}),
  \urlprefix\url{https://doi.org/10.1038/nnano.2013.277}.

\bibitem[{\citenamefont{Kim et~al.}(2008)\citenamefont{Kim, Jin, Moon, Kim,
  Park, Leem, Yu, Noh, Kim, Oh et~al.}}]{BJ2008Iridate}
\bibinfo{author}{\bibfnamefont{B.~J.} \bibnamefont{Kim}},
  \bibinfo{author}{\bibfnamefont{H.}~\bibnamefont{Jin}},
  \bibinfo{author}{\bibfnamefont{S.~J.} \bibnamefont{Moon}},
  \bibinfo{author}{\bibfnamefont{J.-Y.} \bibnamefont{Kim}},
  \bibinfo{author}{\bibfnamefont{B.-G.} \bibnamefont{Park}},
  \bibinfo{author}{\bibfnamefont{C.~S.} \bibnamefont{Leem}},
  \bibinfo{author}{\bibfnamefont{J.}~\bibnamefont{Yu}},
  \bibinfo{author}{\bibfnamefont{T.~W.} \bibnamefont{Noh}},
  \bibinfo{author}{\bibfnamefont{C.}~\bibnamefont{Kim}},
  \bibinfo{author}{\bibfnamefont{S.-J.} \bibnamefont{Oh}},
  \bibnamefont{et~al.}, \bibinfo{journal}{Phys. Rev. Lett.}
  \textbf{\bibinfo{volume}{101}}, \bibinfo{pages}{076402}
  (\bibinfo{year}{2008}),
  \urlprefix\url{https://link.aps.org/doi/10.1103/PhysRevLett.101.076402}.

\bibitem[{\citenamefont{Min et~al.}(2019)\citenamefont{Min, Bentmann, Neu, Eck,
  Moser, Figgemeier, \"Unzelmann, Kissner, Lutz, Koch
  et~al.}}]{Chul-Hee2019TaPWeyl}
\bibinfo{author}{\bibfnamefont{C.-H.} \bibnamefont{Min}},
  \bibinfo{author}{\bibfnamefont{H.}~\bibnamefont{Bentmann}},
  \bibinfo{author}{\bibfnamefont{J.~N.} \bibnamefont{Neu}},
  \bibinfo{author}{\bibfnamefont{P.}~\bibnamefont{Eck}},
  \bibinfo{author}{\bibfnamefont{S.}~\bibnamefont{Moser}},
  \bibinfo{author}{\bibfnamefont{T.}~\bibnamefont{Figgemeier}},
  \bibinfo{author}{\bibfnamefont{M.}~\bibnamefont{\"Unzelmann}},
  \bibinfo{author}{\bibfnamefont{K.}~\bibnamefont{Kissner}},
  \bibinfo{author}{\bibfnamefont{P.}~\bibnamefont{Lutz}},
  \bibinfo{author}{\bibfnamefont{R.~J.} \bibnamefont{Koch}},
  \bibnamefont{et~al.}, \bibinfo{journal}{Phys. Rev. Lett.}
  \textbf{\bibinfo{volume}{122}}, \bibinfo{pages}{116402}
  (\bibinfo{year}{2019}),
  \urlprefix\url{https://link.aps.org/doi/10.1103/PhysRevLett.122.116402}.

\bibitem[{\citenamefont{Denlinger et~al.}(2022)\citenamefont{Denlinger, Kang,
  Dudy, Allen, Kim, Shim, Haule, Sarrao, Butch, and Maple}}]{Denlinger2022}
\bibinfo{author}{\bibfnamefont{J.~D.} \bibnamefont{Denlinger}},
  \bibinfo{author}{\bibfnamefont{J.-S.} \bibnamefont{Kang}},
  \bibinfo{author}{\bibfnamefont{L.}~\bibnamefont{Dudy}},
  \bibinfo{author}{\bibfnamefont{J.~W.} \bibnamefont{Allen}},
  \bibinfo{author}{\bibfnamefont{K.}~\bibnamefont{Kim}},
  \bibinfo{author}{\bibfnamefont{J.-H.} \bibnamefont{Shim}},
  \bibinfo{author}{\bibfnamefont{K.}~\bibnamefont{Haule}},
  \bibinfo{author}{\bibfnamefont{J.~L.} \bibnamefont{Sarrao}},
  \bibinfo{author}{\bibfnamefont{N.~P.} \bibnamefont{Butch}}, \bibnamefont{and}
  \bibinfo{author}{\bibfnamefont{M.~B.} \bibnamefont{Maple}},
  \bibinfo{journal}{Electronic Structure} \textbf{\bibinfo{volume}{4}},
  \bibinfo{pages}{013001} (\bibinfo{year}{2022}),
  \urlprefix\url{https://doi.org/10.1088/2516-1075/ac4315}.

\bibitem[{\citenamefont{Alet and Laflorencie}(2018)}]{ALET2018498}
\bibinfo{author}{\bibfnamefont{F.}~\bibnamefont{Alet}} \bibnamefont{and}
  \bibinfo{author}{\bibfnamefont{N.}~\bibnamefont{Laflorencie}},
  \bibinfo{journal}{Comptes Rendus Physique} \textbf{\bibinfo{volume}{19}},
  \bibinfo{pages}{498} (\bibinfo{year}{2018}), ISSN \bibinfo{issn}{1631-0705},
  \bibinfo{note}{quantum simulation / Simulation quantique},
  \urlprefix\url{https://www.sciencedirect.com/science/article/pii/S163107051830032X}.

\bibitem[{\citenamefont{Abanin et~al.}(2019)\citenamefont{Abanin, Altman,
  Bloch, and Serbyn}}]{RevModPhys.91.021001}
\bibinfo{author}{\bibfnamefont{D.~A.} \bibnamefont{Abanin}},
  \bibinfo{author}{\bibfnamefont{E.}~\bibnamefont{Altman}},
  \bibinfo{author}{\bibfnamefont{I.}~\bibnamefont{Bloch}}, \bibnamefont{and}
  \bibinfo{author}{\bibfnamefont{M.}~\bibnamefont{Serbyn}},
  \bibinfo{journal}{Rev. Mod. Phys.} \textbf{\bibinfo{volume}{91}},
  \bibinfo{pages}{021001} (\bibinfo{year}{2019}),
  \urlprefix\url{https://link.aps.org/doi/10.1103/RevModPhys.91.021001}.

\bibitem[{\citenamefont{Chamon and Mudry}(2001)}]{Chamon-dwave}
\bibinfo{author}{\bibfnamefont{C.}~\bibnamefont{Chamon}} \bibnamefont{and}
  \bibinfo{author}{\bibfnamefont{C.}~\bibnamefont{Mudry}},
  \bibinfo{journal}{Phys. Rev. B} \textbf{\bibinfo{volume}{63}},
  \bibinfo{pages}{100503} (\bibinfo{year}{2001}),
  \urlprefix\url{https://link.aps.org/doi/10.1103/PhysRevB.63.100503}.

\bibitem[{\citenamefont{Basko et~al.}(2006)\citenamefont{Basko, Aleiner, and
  Altshuler}}]{BASKO-MBL}
\bibinfo{author}{\bibfnamefont{D.}~\bibnamefont{Basko}},
  \bibinfo{author}{\bibfnamefont{I.}~\bibnamefont{Aleiner}}, \bibnamefont{and}
  \bibinfo{author}{\bibfnamefont{B.}~\bibnamefont{Altshuler}},
  \bibinfo{journal}{Annals of Physics} \textbf{\bibinfo{volume}{321}},
  \bibinfo{pages}{1126} (\bibinfo{year}{2006}), ISSN \bibinfo{issn}{0003-4916},
  \urlprefix\url{https://www.sciencedirect.com/science/article/pii/S0003491605002630}.

\bibitem[{\citenamefont{Gornyi et~al.}(2005)\citenamefont{Gornyi, Mirlin, and
  Polyakov}}]{Gornyi-MBL}
\bibinfo{author}{\bibfnamefont{I.~V.} \bibnamefont{Gornyi}},
  \bibinfo{author}{\bibfnamefont{A.~D.} \bibnamefont{Mirlin}},
  \bibnamefont{and} \bibinfo{author}{\bibfnamefont{D.~G.}
  \bibnamefont{Polyakov}}, \bibinfo{journal}{Phys. Rev. Lett.}
  \textbf{\bibinfo{volume}{95}}, \bibinfo{pages}{206603}
  (\bibinfo{year}{2005}),
  \urlprefix\url{https://link.aps.org/doi/10.1103/PhysRevLett.95.206603}.

\bibitem[{\citenamefont{Schreiber et~al.}(2015)\citenamefont{Schreiber,
  Hodgman, Bordia, Lüschen, Fischer, Vosk, Altman, Schneider, and
  Bloch}}]{schreiber-observation}
\bibinfo{author}{\bibfnamefont{M.}~\bibnamefont{Schreiber}},
  \bibinfo{author}{\bibfnamefont{S.~S.} \bibnamefont{Hodgman}},
  \bibinfo{author}{\bibfnamefont{P.}~\bibnamefont{Bordia}},
  \bibinfo{author}{\bibfnamefont{H.~P.} \bibnamefont{Lüschen}},
  \bibinfo{author}{\bibfnamefont{M.~H.} \bibnamefont{Fischer}},
  \bibinfo{author}{\bibfnamefont{R.}~\bibnamefont{Vosk}},
  \bibinfo{author}{\bibfnamefont{E.}~\bibnamefont{Altman}},
  \bibinfo{author}{\bibfnamefont{U.}~\bibnamefont{Schneider}},
  \bibnamefont{and} \bibinfo{author}{\bibfnamefont{I.}~\bibnamefont{Bloch}},
  \bibinfo{journal}{Science} \textbf{\bibinfo{volume}{349}},
  \bibinfo{pages}{842} (\bibinfo{year}{2015}),
  \eprint{https://www.science.org/doi/pdf/10.1126/science.aaa7432},
  \urlprefix\url{https://www.science.org/doi/abs/10.1126/science.aaa7432}.

\bibitem[{\citenamefont{yoon Choi et~al.}(2016)\citenamefont{yoon Choi, Hild,
  Zeiher, Schauß, Rubio-Abadal, Yefsah, Khemani, Huse, Bloch, and
  Gross}}]{Gross-exploring}
\bibinfo{author}{\bibfnamefont{J.}~\bibnamefont{yoon Choi}},
  \bibinfo{author}{\bibfnamefont{S.}~\bibnamefont{Hild}},
  \bibinfo{author}{\bibfnamefont{J.}~\bibnamefont{Zeiher}},
  \bibinfo{author}{\bibfnamefont{P.}~\bibnamefont{Schauß}},
  \bibinfo{author}{\bibfnamefont{A.}~\bibnamefont{Rubio-Abadal}},
  \bibinfo{author}{\bibfnamefont{T.}~\bibnamefont{Yefsah}},
  \bibinfo{author}{\bibfnamefont{V.}~\bibnamefont{Khemani}},
  \bibinfo{author}{\bibfnamefont{D.~A.} \bibnamefont{Huse}},
  \bibinfo{author}{\bibfnamefont{I.}~\bibnamefont{Bloch}}, \bibnamefont{and}
  \bibinfo{author}{\bibfnamefont{C.}~\bibnamefont{Gross}},
  \bibinfo{journal}{Science} \textbf{\bibinfo{volume}{352}},
  \bibinfo{pages}{1547} (\bibinfo{year}{2016}),
  \eprint{https://www.science.org/doi/pdf/10.1126/science.aaf8834},
  \urlprefix\url{https://www.science.org/doi/abs/10.1126/science.aaf8834}.

\bibitem[{\citenamefont{L\"uschen et~al.}(2017)\citenamefont{L\"uschen, Bordia,
  Scherg, Alet, Altman, Schneider, and Bloch}}]{Luschen-slowdown}
\bibinfo{author}{\bibfnamefont{H.~P.} \bibnamefont{L\"uschen}},
  \bibinfo{author}{\bibfnamefont{P.}~\bibnamefont{Bordia}},
  \bibinfo{author}{\bibfnamefont{S.}~\bibnamefont{Scherg}},
  \bibinfo{author}{\bibfnamefont{F.}~\bibnamefont{Alet}},
  \bibinfo{author}{\bibfnamefont{E.}~\bibnamefont{Altman}},
  \bibinfo{author}{\bibfnamefont{U.}~\bibnamefont{Schneider}},
  \bibnamefont{and} \bibinfo{author}{\bibfnamefont{I.}~\bibnamefont{Bloch}},
  \bibinfo{journal}{Phys. Rev. Lett.} \textbf{\bibinfo{volume}{119}},
  \bibinfo{pages}{260401} (\bibinfo{year}{2017}),
  \urlprefix\url{https://link.aps.org/doi/10.1103/PhysRevLett.119.260401}.

\bibitem[{\citenamefont{Hwang and Das~Sarma}(2008)}]{Hwang-Born}
\bibinfo{author}{\bibfnamefont{E.~H.} \bibnamefont{Hwang}} \bibnamefont{and}
  \bibinfo{author}{\bibfnamefont{S.}~\bibnamefont{Das~Sarma}},
  \bibinfo{journal}{Phys. Rev. B} \textbf{\bibinfo{volume}{77}},
  \bibinfo{pages}{081412} (\bibinfo{year}{2008}),
  \urlprefix\url{https://link.aps.org/doi/10.1103/PhysRevB.77.081412}.

\bibitem[{\citenamefont{Durham}(1981)}]{Durham_1981}
\bibinfo{author}{\bibfnamefont{P.~J.} \bibnamefont{Durham}},
  \bibinfo{journal}{Journal of Physics F: Metal Physics}
  \textbf{\bibinfo{volume}{11}}, \bibinfo{pages}{2475} (\bibinfo{year}{1981}),
  \urlprefix\url{https://doi.org/10.1088/0305-4608/11/11/027}.

\bibitem[{\citenamefont{Bansil and Pessa}(1983)}]{Bansil_1983}
\bibinfo{author}{\bibfnamefont{A.}~\bibnamefont{Bansil}} \bibnamefont{and}
  \bibinfo{author}{\bibfnamefont{M.}~\bibnamefont{Pessa}},
  \bibinfo{journal}{Physica Scripta} \textbf{\bibinfo{volume}{T4}},
  \bibinfo{pages}{52} (\bibinfo{year}{1983}),
  \urlprefix\url{https://doi.org/10.1088/0031-8949/1983/t4/009}.

\bibitem[{\citenamefont{Starykh}(2015)}]{Starykh2015}
\bibinfo{author}{\bibfnamefont{O.~A.} \bibnamefont{Starykh}},
  \bibinfo{journal}{Reports on Progress in Physics}
  \textbf{\bibinfo{volume}{78}}, \bibinfo{pages}{052502}
  (\bibinfo{year}{2015}),
  \urlprefix\url{https://doi.org/10.1088/0034-4885/78/5/052502}.

\bibitem[{\citenamefont{Lecheminant and Orignac}(2004)}]{Lecheminat-Z3}
\bibinfo{author}{\bibfnamefont{P.}~\bibnamefont{Lecheminant}} \bibnamefont{and}
  \bibinfo{author}{\bibfnamefont{E.}~\bibnamefont{Orignac}},
  \bibinfo{journal}{Phys. Rev. B} \textbf{\bibinfo{volume}{69}},
  \bibinfo{pages}{174409} (\bibinfo{year}{2004}),
  \urlprefix\url{https://link.aps.org/doi/10.1103/PhysRevB.69.174409}.

\bibitem[{\citenamefont{Essler and Tsvelik}(2002)}]{Essler-1DMott}
\bibinfo{author}{\bibfnamefont{F.~H.~L.} \bibnamefont{Essler}}
  \bibnamefont{and} \bibinfo{author}{\bibfnamefont{A.~M.}
  \bibnamefont{Tsvelik}}, \bibinfo{journal}{Phys. Rev. B}
  \textbf{\bibinfo{volume}{65}}, \bibinfo{pages}{115117}
  (\bibinfo{year}{2002}),
  \urlprefix\url{https://link.aps.org/doi/10.1103/PhysRevB.65.115117}.

\bibitem[{\citenamefont{Starykh et~al.}(2010)\citenamefont{Starykh, Katsura,
  and Balents}}]{Starykh2010}
\bibinfo{author}{\bibfnamefont{O.~A.} \bibnamefont{Starykh}},
  \bibinfo{author}{\bibfnamefont{H.}~\bibnamefont{Katsura}}, \bibnamefont{and}
  \bibinfo{author}{\bibfnamefont{L.}~\bibnamefont{Balents}},
  \bibinfo{journal}{Phys. Rev. B} \textbf{\bibinfo{volume}{82}},
  \bibinfo{pages}{014421} (\bibinfo{year}{2010}),
  \urlprefix\url{https://link.aps.org/doi/10.1103/PhysRevB.82.014421}.

\bibitem[{\citenamefont{Starykh and Balents}(2007)}]{Starykh2007}
\bibinfo{author}{\bibfnamefont{O.~A.} \bibnamefont{Starykh}} \bibnamefont{and}
  \bibinfo{author}{\bibfnamefont{L.}~\bibnamefont{Balents}},
  \bibinfo{journal}{Phys. Rev. Lett.} \textbf{\bibinfo{volume}{98}},
  \bibinfo{pages}{077205} (\bibinfo{year}{2007}),
  \urlprefix\url{https://link.aps.org/doi/10.1103/PhysRevLett.98.077205}.

\bibitem[{\citenamefont{Berthod}(2018)}]{Berthod2018}
\bibinfo{author}{\bibfnamefont{C.}~\bibnamefont{Berthod}},
  \emph{\bibinfo{title}{Spectroscopic Probes of Quantum Matter}}, 2053-2563
  (\bibinfo{publisher}{IOP Publishing}, \bibinfo{year}{2018}), ISBN
  \bibinfo{isbn}{978-0-7503-1741-2},
  \urlprefix\url{https://dx.doi.org/10.1088/978-0-7503-1741-2}.

\bibitem[{\citenamefont{Almbladh}(2006)}]{Almbladh_2006}
\bibinfo{author}{\bibfnamefont{C.-O.} \bibnamefont{Almbladh}},
  \bibinfo{journal}{Journal of Physics: Conference Series}
  \textbf{\bibinfo{volume}{35}}, \bibinfo{pages}{127} (\bibinfo{year}{2006}),
  \urlprefix\url{https://doi.org/10.1088/1742-6596/35/1/011}.

\bibitem[{\citenamefont{Bardyszewski and Hedin}(1985)}]{Bardyszewski_1985}
\bibinfo{author}{\bibfnamefont{W.}~\bibnamefont{Bardyszewski}}
  \bibnamefont{and} \bibinfo{author}{\bibfnamefont{L.}~\bibnamefont{Hedin}},
  \bibinfo{journal}{Physica Scripta} \textbf{\bibinfo{volume}{32}},
  \bibinfo{pages}{439} (\bibinfo{year}{1985}),
  \urlprefix\url{https://doi.org/10.1088/0031-8949/32/4/033}.

\bibitem[{\citenamefont{Yasuhara et~al.}(1999)\citenamefont{Yasuhara,
  Yoshinaga, and Higuchi}}]{Yasuchara-weird-S}
\bibinfo{author}{\bibfnamefont{H.}~\bibnamefont{Yasuhara}},
  \bibinfo{author}{\bibfnamefont{S.}~\bibnamefont{Yoshinaga}},
  \bibnamefont{and} \bibinfo{author}{\bibfnamefont{M.}~\bibnamefont{Higuchi}},
  \bibinfo{journal}{Phys. Rev. Lett.} \textbf{\bibinfo{volume}{83}},
  \bibinfo{pages}{3250} (\bibinfo{year}{1999}),
  \urlprefix\url{https://link.aps.org/doi/10.1103/PhysRevLett.83.3250}.

\bibitem[{\citenamefont{McMullen et~al.}(1976)\citenamefont{McMullen,
  Bergersen, and Jena}}]{McMullen_1976}
\bibinfo{author}{\bibfnamefont{T.}~\bibnamefont{McMullen}},
  \bibinfo{author}{\bibfnamefont{B.}~\bibnamefont{Bergersen}},
  \bibnamefont{and} \bibinfo{author}{\bibfnamefont{P.}~\bibnamefont{Jena}},
  \bibinfo{journal}{Journal of Physics C: Solid State Physics}
  \textbf{\bibinfo{volume}{9}}, \bibinfo{pages}{975} (\bibinfo{year}{1976}),
  \urlprefix\url{https://doi.org/10.1088/0022-3719/9/6/016}.

\bibitem[{\citenamefont{Witten and
  Sander}(1981{\natexlab{a}})}]{Witten-Brown-tree}
\bibinfo{author}{\bibfnamefont{T.~A.} \bibnamefont{Witten}} \bibnamefont{and}
  \bibinfo{author}{\bibfnamefont{L.~M.} \bibnamefont{Sander}},
  \bibinfo{journal}{Phys. Rev. Lett.} \textbf{\bibinfo{volume}{47}},
  \bibinfo{pages}{1400} (\bibinfo{year}{1981}{\natexlab{a}}),
  \urlprefix\url{https://link.aps.org/doi/10.1103/PhysRevLett.47.1400}.

\bibitem[{\citenamefont{Eggert and Affleck}(1992)}]{EggertAffleck}
\bibinfo{author}{\bibfnamefont{S.}~\bibnamefont{Eggert}} \bibnamefont{and}
  \bibinfo{author}{\bibfnamefont{I.}~\bibnamefont{Affleck}},
  \bibinfo{journal}{Phys. Rev. B} \textbf{\bibinfo{volume}{46}},
  \bibinfo{pages}{10866} (\bibinfo{year}{1992}),
  \urlprefix\url{https://link.aps.org/doi/10.1103/PhysRevB.46.10866}.

\bibitem[{\citenamefont{Braun et~al.}(2013)\citenamefont{Braun, Min\'ar,
  Mankovsky, Strocov, Brookes, Plucinski, Schneider, Fadley, and
  Ebert}}]{Braun-phononARPES}
\bibinfo{author}{\bibfnamefont{J.}~\bibnamefont{Braun}},
  \bibinfo{author}{\bibfnamefont{J.}~\bibnamefont{Min\'ar}},
  \bibinfo{author}{\bibfnamefont{S.}~\bibnamefont{Mankovsky}},
  \bibinfo{author}{\bibfnamefont{V.~N.} \bibnamefont{Strocov}},
  \bibinfo{author}{\bibfnamefont{N.~B.} \bibnamefont{Brookes}},
  \bibinfo{author}{\bibfnamefont{L.}~\bibnamefont{Plucinski}},
  \bibinfo{author}{\bibfnamefont{C.~M.} \bibnamefont{Schneider}},
  \bibinfo{author}{\bibfnamefont{C.~S.} \bibnamefont{Fadley}},
  \bibnamefont{and} \bibinfo{author}{\bibfnamefont{H.}~\bibnamefont{Ebert}},
  \bibinfo{journal}{Phys. Rev. B} \textbf{\bibinfo{volume}{88}},
  \bibinfo{pages}{205409} (\bibinfo{year}{2013}),
  \urlprefix\url{https://link.aps.org/doi/10.1103/PhysRevB.88.205409}.

\bibitem[{\citenamefont{Feder}(1981)}]{Feder_1981}
\bibinfo{author}{\bibfnamefont{R.}~\bibnamefont{Feder}},
  \bibinfo{journal}{Journal of Physics C: Solid State Physics}
  \textbf{\bibinfo{volume}{14}}, \bibinfo{pages}{2049} (\bibinfo{year}{1981}),
  \urlprefix\url{https://doi.org/10.1088/0022-3719/14/15/006}.

\bibitem[{\citenamefont{Witten and Sander}(1981{\natexlab{b}})}]{DLA-orig}
\bibinfo{author}{\bibfnamefont{T.~A.} \bibnamefont{Witten}} \bibnamefont{and}
  \bibinfo{author}{\bibfnamefont{L.~M.} \bibnamefont{Sander}},
  \bibinfo{journal}{Phys. Rev. Lett.} \textbf{\bibinfo{volume}{47}},
  \bibinfo{pages}{1400} (\bibinfo{year}{1981}{\natexlab{b}}),
  \urlprefix\url{https://link.aps.org/doi/10.1103/PhysRevLett.47.1400}.

\bibitem[{\citenamefont{Wu et~al.}(1976)\citenamefont{Wu, McCoy, Tracy, and
  Barouch}}]{Wu1976Ising}
\bibinfo{author}{\bibfnamefont{T.~T.} \bibnamefont{Wu}},
  \bibinfo{author}{\bibfnamefont{B.~M.} \bibnamefont{McCoy}},
  \bibinfo{author}{\bibfnamefont{C.~A.} \bibnamefont{Tracy}}, \bibnamefont{and}
  \bibinfo{author}{\bibfnamefont{E.}~\bibnamefont{Barouch}},
  \bibinfo{journal}{Phys. Rev. B} \textbf{\bibinfo{volume}{13}},
  \bibinfo{pages}{316} (\bibinfo{year}{1976}),
  \urlprefix\url{https://link.aps.org/doi/10.1103/PhysRevB.13.316}.

\bibitem[{\citenamefont{Gursoy et~al.}(2012)\citenamefont{Gursoy, Plauschinn,
  Stoof, and Vandoren}}]{GursoyARPES}
\bibinfo{author}{\bibfnamefont{U.}~\bibnamefont{Gursoy}},
  \bibinfo{author}{\bibfnamefont{E.}~\bibnamefont{Plauschinn}},
  \bibinfo{author}{\bibfnamefont{H.}~\bibnamefont{Stoof}}, \bibnamefont{and}
  \bibinfo{author}{\bibfnamefont{S.}~\bibnamefont{Vandoren}},
  \bibinfo{journal}{Journal of High Energy Physics}
  \textbf{\bibinfo{volume}{2012}}, \bibinfo{pages}{18} (\bibinfo{year}{2012}),
  ISSN \bibinfo{issn}{1029-8479},
  \urlprefix\url{https://doi.org/10.1007/JHEP05(2012)018}.

\end{thebibliography}
\end{document}